# An image of the heteroatom influence in acyclic polyenes

## Viktorija Gineityte

Institute of Theoretical Physics and Astronomy, Vilnius University, Sauletekio al. 3, LT- 10257 Vilnius, Lithuania **E-mail:** viktorija.gineityte@tfai.vu.lt

**Abstract**
The study is devoted to development of the theory of the heteroatom influence in acyclic polyenes from a local perspective. The charge redistribution due to introduction of heteroatom(s) (X) is decomposed into additive and meaningful contributions undergoing an extinction with growth of the substructure embraced. Each of these contributions, in turn, is traced back to specific additional interactions of bond orbitals (BOs) of C=C bonds of the parent hydrocarbon. The expansion starts with local and transferable dipoles of individual X=C (or C=X) bonds that are proportional to direct interactions of BOs within respective parent C=C bonds. Thereupon, populations transferred between first-neighbouring double bonds follow along with parallel and anti-parallel secondary (induced) dipoles of the latter. These contributions are determined by an interplay of two indirect interbond interactions of BOs and consequently depend on positions of heteroatom(s) inside the bonds concerned.

## Introduction

Organic molecules with heteroatoms are traditionally treated in chemistry as derivatives of respective parent hydrocarbons (see e.g. [1, 2]). This actually means that properties of the former are believed to depend somehow on the constitution of the latter. Moreover, such a dependence usually is expected to be more or less similar for parent hydrocarbons of the same class, e.g. saturated, conjugated, etc. On this basis, the heteroatom influence is regarded as a certain single phenomenon embracing the whole class of substituted compounds concerned. The above-outlined perspective, in turn, stimulates the search for the relevant general theoretical approaches. In this area, simple models of structures of a wide scope of applicability are preferable over exact numerical calculations of individual compounds.

Development of quantum-chemical approaches to the heteroatom influence in conjugated hydrocarbons appearently started with the well-known classical perturbation theory of Coulson and Longuet-Higgins (abbreviated below as CLHPT) [3, 4], where heteroatoms have been represented by small alterations in Coulomb integrals of the standard Hückel model [5-7] and self- and mutual polarizabilities of atoms of the parent hydrocarbon played the role of the principal quantities. Analysis of expressions for the latter resulted into relations of the above-anticipated type, i.e. between the structure of the parent hydrocarbon and properties of the substituted compound. In our context, the rule of the alternating polarity [4] for alternant hydrocarbons deserves mention as an achievement of this theory. Further, the subsequently revealed proportionality between the Hückel type Hamiltonian matrix of any conjugated hydrocarbon and the adjacency matrix (AM) of the graph of its carbon backbone [8] gave rise to an additional employment of powerful graph-theoretical methods in studies of the heteroatom influence. The derivatives of conjugated hydrocarbons have been modelled here by somewhat generalized graphs [9-12] containing self-loops (weighted vertices) at the site(s) of heteroatom(s) without any small-value condition for the weight. The relevant theory then enabled one to go beyond the perturbational treatment of heteroatom(s) and thereby embraced the CHLPT as a particular case. The underlying technique, however, became more involved and cumbersome, especially for systems with more than one heteroatom [12]. Nevertheless, properties of alternative derivatives (e.g. their relative stabilities) proved to be determined by descriptors of the standard graph of the respective parent hydrocarbon, viz. by the so-called topological index [9] and/or the Dewar index [10] at the site(s) of substitution in accordance with the above-discussed chemical tradition. Finally, a fortunate combination of the CLHPT with a definite graph-theoretical technique [13] resulted into rules representing the extinction rates of mutual polarizabilities ($\pi_{rs}$) with increasing distances between the $r$th and $s$th carbon atoms of an acyclic polyene. These rules offered a substantiation of some conlusions of the "curly arrow chemistry" [14-16] regarding specific derivatives of these hydrocarbons.

Another important (and not yet achieved) goal of the theory of the heteroatom influence consists in gaining a knowledge of how and why the phenomenon itself arises, i.e. in constructing an image of the latter. In particular, this task embraces the revealing of specific interorbital interactions behind the charge redistribution due to introduction of heteroatom(s), as well as of the most important participating substructures (fragments) of the system. In the present study, an attempt is undertaken just in this direction.

Creation of any mental image usually is based on decomposition of the phenomenon under study into certain elementary components followed by selection of the most important ones. So far as chemistry is concerned, additive and essentially local components always are preferable (cf. the well-known local image of conjugation as a weak interaction of quasi-transferable unsaturated functional groups [1, 2, 17]). Meanwhile, both the molecular graph and the standard Hückel model are orientated mostly towards representing the constitution of the carbon backbone as a whole. In particular, self- and mutual polarizabilities are expressible via coefficients of respective delocalized (canonical) molecular orbitals (MOs) [3, 18, 19] and thereby belong to global characteristics of the parent hydrocarbon. The same refers also to the above-mentioned indices of Refs.[9, 10].

With this in mind, a question arises whether the molecular graph itself is the most appropriate model for achieving our end. In respect of feasibility of other options, however, a clear distinction should be made between (poly)cyclic (aromatic) and acyclic hydrocarbons. Indeed, employment of the molecular graph as a model is based on an assumption that carbon-carbon bonds are uniform (or almost uniform) over the whole backbone of the conjugated hydrocarbon concerned. Such a requirement is more or less met for (poly)cyclic (aromatic) hydrocarbons, particularly for benzenoids (see e.g. [20]). Meanwhile, the analogous bond lengths are far from being uniform over chains of acyclic polyenes, especially of small- and medium-sized ones. The point is that formally-double (C=C) bonds still remain significantly shorter here as compared to formally-single (C–C) ones, although these are somewhat lengthened and shortened, respectively, vs. their standard values [21-24] (cf. the so-called bond length alternation (BLA) [22, 24]). Consequently, the molecular graph and/or its AM is merely one of two possible approximate models of acyclic polyenes. Another alternative consists in consideration of an acyclic $\pi$-electron system as a set of weakly-interacting formally-double (C=C) bonds, where the formally-single (C–C) bonds play the role of the interaction and may be consequently regarded as a perturbation [25, 26]. Such a perturbational treatment of the carbon backbone itself evidently is in line with the above-mentioned local image of conjugation. It also deserves recalling here that the results of any perturbation theory are characterized by a relatively high degree of additivity [27]. Thus, we expect good prospects of just this alternative model for constructing an image of the heteroatom influence. That is why we confine ourselves to the case of acyclic polyenes as parent systems in the present study.

The development of the perturbational perspective on acyclic $\pi$-electron systems has a long history starting with applications of the simple PMO theory [27] and ending with approaches based on a perturbative treatment of the configuration interaction [28, 29]. In the latter case, the most important configuration (the zero-order many-electron wave function) coincides with the determinant built up of localized double-occupied orbitals of formally-double (C=C) bonds, whereas contributions of the remaining (excited) configurations are considered as a perturbation. An analogous methodology seems to underly also the recent valence bond self-consistent field calculations of polyenes [30, 31], where mixing of separate Rumer structures was found to follow the standard rules of the perturbation theory.

It is noteworthy, however, that the above-overviewed perturbational studies are devoted mainly to analysis of expressions for total $\pi$-electron energies of acyclic hydrocarbons (polyenes) and/or their many-electron wave functions. Meanwhile, charge redistribution over the carbon backbone is the principal manifestation of the heteroatom influence and, consequently, expressions for one-electron density matrices and/or charge-bond order (CBO) matrices of substituted compounds are required in our case. To achieve this end, the direct method of derivation of these matrices [32-34] is especially attractive that is based on solution of the so-called commutation equation [35] in the form of power series. The main reasons for this option are as follows: First, the above-mentioned solution is carried out without any reference to delocalized (canonical) MOs and, consequently, local relations (if any) are preserved between individual elements of the CBO matrix, on the one hand, and those of the relevant Hamiltonian matrix, on the other hand, i.e. between the charge redistribution and the local structure. Second, the commutation equation is solvable for entire classes of molecules under certain conditions. In particular, representability of initial Hamiltonian matrices of all acyclic polyenes in a common form via certain zero and first order submatrices (blocks) enabled us previously to derive an analogous common expression for the relevant CBO matrix [34, 36, 37]. Under an additional assumption that all C–C bonds and all C=C ones are uniform [25, 26], the relevant power series contains a single perturbational parameter ($\gamma$) coinciding with the ratio of the resonance integral of the former to that of of the latter. At the same time, separate members of the series proved to be representable via local (direct and indirect) interactions between orbitals of double (C=C) bonds [34]. The latter aspect seems to be especially prospective when looking for an image of the heteroatom influence.

Thus, we are now about to extend the above-described achievements to acyclic polyenes with any number of heteroatoms irrespective of their positions inside the parent hydrocarbon. To this end, we will introduce an additional perturbational parameter ($\alpha$) into the former model of the parent polyene [25, 26] that coincides with the averaged alteration in the relevant Coulomb integral(s) of carbon atom(s) after replacing them by heteroatom(s). The ultimate aim of the study then consists in analysis of the relevant dual series for the common CBO matrix of polyenes with heteroatom(s).

## Results and discussion

*General properties of the charge redistribution*

Let us start with the perturbational model of the parent polyenes. Let us consider a certain polyene ($C_{2N}H_{2N+2}$) of arbitrary constitution (linear or branched) containing $N$ formally-double (C=C) bonds. We will confine ourselves here to small- and medium-sized acyclic chains. At the Hückel level [5-7], the $\pi$-electron system of this molecule is representable by the $2N$-dimensional basis of $2p_z$ atomic orbitals (AOs) of carbon atoms {$\chi$}. Let these AOs be characterized by uniform Coulomb integrals coinciding with the energy reference point and consequently taking a zero value as usual. Resonance integrals between AOs inside formally-double bonds also are assumed to be uniform and are chosen as the (negative) energy unit in addition. Meanwhile, the analogous uniform integrals of formally-single (C–C) bonds will be denoted by $\gamma$ and supposed to be a small (first order) term vs. the above-specified energy unit. Finally, resonance integrals for pairs of more remote AOs will be ignored [5-7]. Under the above-enumerated conditions, our polyene may be considered as a set of $N$ initially-isolated C=C bonds undergoing a simple perturbation associated with C–C bonds and characterized by a single positive dimensionless parameter $\gamma$ (the latter actually coinciding with the ratio of the resonance integral of C–C bonds to that of C=C bonds). Furthermore, our polyene belongs to alternant hydrocarbons [5-8, 18, 19, 27] under the above-described approximations. Consequently, the relevant basis set $\{\chi\}$ always is divisible into two $N$-dimensional subsets $\{\chi^1\}$ and $\{\chi^2\}$ so that pairs of orbitals belonging to chemical bonds (C=C or C–C) find themselves in different subsets [8]. This implies non-zero resonance parameters (1 and $\gamma$, respectively) to take place in the off-diagonal (intersubset) blocks of our initial Hamiltonian matrix $H$ [25, 26] as exhibited below. Given that the AOs are enumerated in such a way that orbitals belonging to the same C=C bond (say, to the Ith one) acquire coupled numbers $i$ and $N+i$, resonance parameters of these strong bonds take the diagonal positions in the intersubset blocks of the matrix $H$.

Let us assume now that a certain number of heteroatoms (X) is introduced into the above-described parent polyene, the relevant electronegativities being slightly different from that of carbon atoms. Let the respective averaged alteration in the Coulomb parameters be denoted by $\alpha$ and supposed to be a first order term vs. the same energy unit. Obviously, a positive parameter $\alpha$ will correspond to the most common case of more electronegative heteroatoms. This implies that our zero-order system (i.e. the set of $N$ isolated C=C bonds) now undergoes a dual perturbation characterized by two parameters $\gamma$ and $\alpha$. As a result, the total Hamiltonian matrix of our heteroatom(s)-containing compound ($H$) is representable as a sum of a zero order member ($H_{(0)}$) and of the first order one ($H_{(1)}$). Moreover, both of these terms consist of four $N$x$N$-dimensional submatrices corresponding to subsets $\{\chi^1\}$ and $\{\chi^2\}$ and to their interaction, viz.

$$H = H_{(0)} + H_{(1)} = \begin{vmatrix} 0 & I \\ I & 0 \end{vmatrix} + \begin{vmatrix} \alpha A & \gamma B \\ \gamma B^* & \alpha C \end{vmatrix}, \tag{1}$$

where $I$ here and below stands for the unit (sub)matrix and the superscript $*$ designates the transposed (Hermitian-conjugate) matrix. The zero order member of Eq.(1) ($H_{(0)}$) contains resonance integrals of C=C bonds, whilst the first order one ($H_{(1)}$) embraces analogous parameters of C–C bonds ($\gamma$) along with those of heteroatoms ($\alpha$). Since entire resonance integrals of C=C bonds are included into the zero order matrix $H_{(0)}$, diagonal elements of the submatrix $B$ take zero values (i.e. $B_{ii}=0$). Meanwhile, the relevant off-diagonal elements ($B_{ij}, i \neq j$) coincide with 1 in the positions associated with C–C bonds and vanish otherwise. By contrast, $A$ and $C$ are diagonal (sub)matrices containing unit elements in the position(s) referring to heteroatom(s). Obviously, zero (sub)matrices $A$ and $C$ correspond to the parent polyene. Introduction of a single heteroatom (X) into a certain (e.g. Ith) C=C bond of the latter is alternatively representable either by an element $A_{ii}=1$ or by $C_{ii}=1$, it depends on the number of the AO of the carbon atom under substitution ($i$ or $N+i$). These options are referred to below as formation of an X=C bond and of a C=X one, respectively. Meanwhile, replacement of the Ith C=C bond by an X=X one is described by the only equality $A_{ii}=C_{ii}=1$. It also deserves emphasizing here that the particular constitution of (sub)matrices $A$, $B$ and $C$ is not specified at the present stage of our discussion. Thus, the matrix $H$ of Eq.(1) actually represents the whole class of polyenes with heteroatom(s).

Let us now turn to derivation of the relevant CBO matrix $P$. It deserves an immediate mention that the above-introduced subsets of AOs $\{\chi^1\}$ and $\{\chi^2\}$ are characterized by a zero intersubset energy gap. Consequently, perturbative approaches cannot be straightforwardly applied to the matrix $H$ of Eq.(1) and a certain transformation of the basis set is required. The simplest option here consists in turning to the basis of bond orbitals (BOs) of C=C bonds of the parent polyene. Let the bonding BO (BBO) of the Ith C=C bond $\phi_{(+)i}$ and its antibonding counterpart (ABO) $\phi_{(-)i}$ be defined as a normalized sum and difference, respectively, of the relevant AOs $\chi_i^1$ and $\chi_{N+i}^2$. Let the subsets of BBOs and of ABOs be correspondingly designated by $\{\phi_{(+)}\}$ and $\{\phi_{(-)}\}$. Passing from the initial basis set $\{\chi\}$ to the new one $\{\phi\}$ is then representable by the following simple unitary matrix

$$U = \frac{1}{\sqrt{2}} \begin{vmatrix} I & I \\ I & -I \end{vmatrix}, \quad (2)$$

which serves to transform the initial Hamiltonian matrix $H$ of Eq.(1). The new Hamiltonian matrix is then as follows

$$H' = H'_{(0)} + H'_{(1)} = \begin{vmatrix} I & 0 \\ 0 & -I \end{vmatrix} + \begin{vmatrix} T & R \\ R* & Q \end{vmatrix}, \quad (3)$$

where the superscript ' serves to distinguish between the present matrix and that of Eq.(1). The $N \times N$-dimensional submatrices $T$ and $Q$ now represent the interactions inside subsets $\{\phi_{(+)}\}$ and $\{\phi_{(-)}\}$, respectively, whereas $R$ embraces intersubset interactions (i.e. resonance parameters between BBOs and ABOs). At the same time, all submatrices of the first order matrix $H'_{(1)}$ consist of zero and first order increments with respect to parameter $\alpha$ denoted below by superscripts (0) and (1), respectively. i.e.

$$T = T^{(0)} + T^{(1)}, \quad Q = Q^{(0)} + Q^{(1)}, \quad R = R^{(0)} + R^{(1)}, \quad (4)$$

where

$$T^{(0)} = -Q^{(0)} = \frac{\gamma}{2}(B* + B), \quad R^{(0)} = -R^{(0)}* = \frac{\gamma}{2}(B* - B),$$

$$T^{(1)} = Q^{(1)} = \frac{\alpha}{2}(A + C), \quad R^{(1)} = R^{(1)}* = \frac{\alpha}{2}(A - C). \quad (5)$$

Obviously, increments $T^{(0)}$, $Q^{(0)}$ and $R^{(0)}$ contain interactions between BOs of the parent polyene. Diagonal elements of these matrices vanish, i.e. $T_{ii}^{(0)} = Q_{ii}^{(0)} = R_{ii}^{(0)} = 0$ (The equality $B_{ii}=0$ should be recalled here). Meanwhile, off-diagonal elements of the same matrices take non-zero values for BOs of first-neighbouring C=C bonds and vanish otherwise [25] (First-neighbouring formally-double bonds are defined as those connected by a formally-single bond). Further, the remaining terms of Eq.(4) (i.e. $T^{(1)}$, $Q^{(1)}$ and $R^{(1)}$) embrace corrections to the above-mentioned interactions due to introduction of heteroatom(s) and are diagonal matrices in contrast to the former. Non-zero elements $T_{ii}^{(1)}$ and $Q_{ii}^{(1)}$ (if any) represent newly-emerging self-interactions of BOs $\phi_{(+)i}$ and $\phi_{(-)i}$, respectively, whereas a significant element $R_{ii}^{(1)}$ indicates an additional resonance parameter to arise between the same BOs that is referred to below as the intrabond one. Particular values of parameters under discussion are easily obtainable after substituting into Eq.(5) the appropriate elements of matrices $A$ and $C$. As a result, self-interactions $T_{ii}^{(1)}(Q_{ii}^{(1)})$ equal to $\alpha/2$ and $\alpha$ for X=C(C=X) and X=X bonds, respectively, whereas non-zero elements $R_{ii}^{(1)}$ actually refer to X=C and C=X bonds only and correspondingly coincide with $\alpha/2$ and $-\alpha/2$. Finally, all increments of Eq.(5) are symmetric matrices except for the only skew-symmetric component $R^{(0)}$.

As is easily seen, the subsets $\{\phi_{(+)}\}$ and $\{\phi_{(-)}\}$ are now characterized by a substantial energy gap ($2I$) vs. the intersubset interaction ($R$). Moreover, comparison of the matrix $H'$ of Eq.(3) to that underlying the direct way of derivation of the CBO matrix (see *Methods*) shows the former to coincide with a particular case of the latter. That is why this derivation is straightforwardly applicable to construct the CBO matrix $P'$ referring to the Hamiltonian matrix $H'$ of Eq.(3). Thereupon, the result may be easily retransformed into the basis of AOs $\{\chi\}$ again by means of the matrix $U$ of Eq.(2) and the CBO matrix under our interest ($P$) ultimately follows.

It deserves emphasizing here, however, that possibility of deriving the very CBO matrix ($P$) is not the only good of passing into the basis of BOs. Indeed, this transformation allows us to introduce some chemically-illustrative concepts (namely, the direct and indirect interorbital interactions discussed below) that are preserved also after returning to the basis $\{\chi\}$ and prove to be efficient in discussions of the charge redistribution. In this connection, let us dwell on a brief overview of the principal features of the intermediate matrix $P'$.

i) The matrix $P'$ is representable in the form of power series, i.e. as a sum of increments $P'_{(k)}$ of an increasing order $k$ with respect to parameters contained within the first order term ($H'_{(1)}$) of the Hamiltonian matrix $H'$ of Eq.(3).

ii) The zero order member of this series ($P'_{(0)}$) and the subsequent corrections ($P'_{(k)}$, $k$=1,2,3…) correspondingly take the following forms

$$P'_{(0)} = 2\begin{vmatrix} I & 0 \\ 0 & 0 \end{vmatrix}, \qquad P'_{(k)} = -2\begin{vmatrix} D^{(+)}_{(k)} & G_{(k)} \\ G^*_{(k)} & -D^{(-)}_{(k)} \end{vmatrix}, \qquad (6)$$

where $P'_{(0)}$ contains the initial occupation numbers of BOs coinding with 2 and 0 for those of subsets $\{\phi_{(+)}\}$ and $\{\phi_{(-)}\}$, respectively. Meanwhile, $P'_{(k)}$, $k$=1,2,3.. consist of certain $N$x$N$-dimensional principal (sub)matrices $G_{(k)}$, $k$=1,2,3.. specified below and of their derivatives $D^{(+)}_{(k)}$ and $D^{(-)}_{(k)}$, $k$=2,3… defined as follows

$$D^{(+)}_{(2)} = G_{(1)}G^*_{(1)}, \qquad D^{(-)}_{(2)} = G^*_{(1)}G_{(1)}, \qquad D^{(+)}_{(3)} = G_{(1)}G^*_{(2)} + G_{(2)}G^*_{(1)},$$
$$D^{(-)}_{(2)} = G^*_{(1)}G_{(2)} + G^*_{(2)}G_{(1)}, etc. \qquad (7)$$

Subscripts of matrices of Eqs.(6) and (7) reflect the above-introduced order parameter $k$, whilst superscripts of matrices $D^{(+)}_{(k)}$ and $D^{(-)}_{(k)}$ point to their links to subsets $\{\phi_{(+)}\}$ and $\{\phi_{(-)}\}$, respectively, in accordance with their positions inside the correction $P'_{(k)}$ of Eq.(6). Finally, the front factor of Eq.(6) (–2) is introduced for convenience.

iii) The principal matrices $G_{(k)}$, $k$=1,2,3.. are representable via entire submatrices $T$, $Q$ and $R$ of Eq.(3), whatever the actual constitution of the latter. Thus, the first order matrix $G_{(1)}$ is proportional to the matrix $R$, whereas those of higher orders ($G_{(2)}$, $G_{(3)}$, etc) meet specific recurrence relations, i.e. the matrix of the $k$th order is expressible via those of lower orders ($k$–1, $k$–2, etc). As a result, matrices $G_{(k)}$, $k$=2,3.. depend on products of submatrices (blocks) $T$, $Q$ and $R$. The overall form of these relations is then as follows

$$G_{(1)} = -\frac{1}{2}R, \qquad G_{(2)} = -\frac{1}{2}(TG_{(1)} - G_{(1)}Q) = \frac{1}{4}(TR - RQ),$$
$$G_{(3)} = -\frac{1}{2}(TG_{(2)} - G_{(2)}Q) - 2D^{(+)}_{(2)}G_{(1)}, \text{ etc.} \qquad (8)$$

iv) Individual elements of matrices $G_{(k)}$, $D^{(+)}_{(k)}$ and $D^{(-)}_{(k)}$ have a clear chemical meaning, viz. these represent definite types of local (direct) and semi-local (indirect) interactions of BOs of C=C bonds of the parent hydrocarbon.

As already mentioned, the above-described matrix $P'$ subsequently undergoes retransformation into the basis of AOs $\{\chi\}$. Due to the simple constitution of the (re)transformation matrix $U$ of Eq.(2), the resulting CBO matrix $P$ contains sums and differences of the same blocks $G_{(k)}$, $D^{(+)}_{(k)}$ and $D^{(-)}_{(k)}$. Accordingly, diagonal elements of the matrix $P$ (i.e. $P_{ii}$) depend on simple combinations of analogous elements of these blocks ($G_{(k)ii}$, $D^{(+)}_{(k)ii}$ and $D^{(-)}_{(k)ii}$, respectively). This implies that occupation numbers of AOs of the Ith double bond ($\chi^1_i$ and $\chi^2_{N+i}$) are determined by those of the relevant two BOs ($\phi_{(+)i}$ and $\phi_{(-)i}$) along with the only bond order between the latter. Thus, let us now consider these three decisive elements of the matrix $P'$ of Eqs.(6)-(8) in a more detail.

Let the occupation numbers of BOs $\phi_{(+)i}$ and $\phi_{(-)i}$ be denoted by $X_i^{(+)}$ and $X_i^{(-)}$, respectively. The relevant expressions then follow from Eq.(6) and take the form of the following power series

$$X_i^{(+)} = 2 + \sum_{k=2}^{\infty} X_{(k)ii}^{(+)}, \qquad X_i^{(-)} = \sum_{k=2}^{\infty} X_{(k)ii}^{(-)}, \qquad (9)$$

where 2 represents the initial (zero order) population of the BBO $\phi_{(+)i}$, whilst contributions of the $k$th order ($X_{(k)ii}^{(+)}$ and $X_{(k)ii}^{(-)}$) are proportional to respective diagonal elements of matrices of Eq.(7), viz.

$$X_{(k)ii}^{(+)} = -2D_{(k)ii}^{(+)}, \qquad X_{(k)ii}^{(-)} = 2D_{(k)ii}^{(-)}. \qquad (10)$$

Expressibility of matrices $\boldsymbol{D}_{(k)}^{(+)}$ and $\boldsymbol{D}_{(k)}^{(-)}$ via products of matrices $\boldsymbol{G}_{(k)}$ of lower orders seen from Eq.(7), in turn, allows the contributions to occupation numbers to be represented as sums of partial increments of separate BOs of the opposite subset, viz.

$$X_{(k)ii}^{(+)} = -\sum_{(-)l} x_{(k)}^{(+)i,(-)l}, \qquad X_{(k)ii}^{(-)} = \sum_{(+)l} x_{(k)}^{(-)i,(+)l}, \qquad (11)$$

where $x_{(k)}^{(+)i,(-)l}$ may be examplified as follows

$$x_{(2)}^{(+)i,(-)l} = 2(G_{(1)il})^2 > 0, \qquad x_{(3)}^{(+)i,(-)l} = 4G_{(1)il}G_{(2)il}, etc \qquad (12)$$

and describe partial populations of the $k$th order transferred inside the pair of orbitals $\phi_{(+)i}$ and $\phi_{(-)l}$ (Note that $x_{(k)}^{(+)i,(-)l}$ coincides with $x_{(k)}^{(-)l,(+)i}$ in accordance with the charge conservation condition). Sums of Eq.(11) over $(-)l$ and over $(+)l$ correspondingly embrace all ABOs and all BBOs of the given system. The minus sign of the first relation of Eq.(11) is introduced for convenience and expresses an aggreement that a positive partial population ($x_{(k)}^{(+)i,(-)l} > 0$) corresponds to the charge transfer from a BBO ($\phi_{(+)i}$) to an ABO ($\phi_{(-)l}$) that is accompanied by reduction of the total occupation number of the former and by an increase of that of the latter. The *a priori* positive sign of the second order increment $x_{(2)}^{(+)i,(-)l}$ (seen from Eq.(12)) also deserves attention here.

As already mentioned, the bond order between the BBO $\phi_{(+)i}$ and the ABO $\phi_{(-)i}$ also is among decisive elements of the matrix $\boldsymbol{P}$. For reasons clarified below, let this bond order be denoted by $d_I$. The expansion concerned is then as follows

$$d_I = \sum_{k=1}^{\infty} d_{(k)I} = -2\sum_{k=1}^{\infty} G_{(k)ii} \qquad (13)$$

and contains diagonal elements ($G_{(k)ii}$) of the principal matrices $\boldsymbol{G}_{(k)}$, $k=1,2,3…$

Let the occupation numbers (populations) of AOs of the Ith double bond ($\chi_i^1$ and $\chi_{N+i}^2$) be denoted by $Y_i^1$ and $Y_{N+i}^2$, respectively. The relevant expressions ultimately take a simple form [34], viz.

$$Y_i^1(Y_{N+i}^2) = 1 + \frac{1}{2}X_I \pm d_I, \qquad (14)$$

where the upper and lower signs of the right-hand side correspondingly refer to AOs $\chi_i^1$ and $\chi_{N+i}^2$. The first term of Eq.(14) (i.e. 1) represents the initial (zero order) population of any AO before perturbation, i.e. inside the set of $N$ isolated C=C bonds. Meanwhile, the second contribution coincides with a half of the total population ($X_I$) lost (acquired) by the Ith double bond due to the overall perturbation. The extra population $X_I$, in turn, equals to the sum of alterations in occupation numbers of BOs $\phi_{(+)i}$ and $\phi_{(-)i}$ and also is accordingly representable in the form of power series, the $k$th order member of which ($X_{(k)I}$, $k=2,3…$) depends on corrections of Eq.(10), viz.

$$X_{(k)I} = X_{(k)ii}^{(+)} + X_{(k)ii}^{(-)} = -2(D_{(k)ii}^{(+)} - D_{(k)ii}^{(-)}). \qquad (15)$$

Moreover, an additional employment of Eq.(11) shows that both $X_I$ and its separate increments $X_{(k)I}$ are expressible as sums of contributions of other double bonds (L). In particular, the contribution of the Lth double bond to $X_{(k)I}$ (denoted below by $X_{(k)I(L)}$) coincides with the difference between the relevant partial transferred populations $x_{(k)}^{(+)l,(-)i}$ and $x_{(k)}^{(+)i,(-)l}$. Accordingly, the increment of the Ith double bond to $X_{(k)L}$ (denoted by $X_{(k)L(I)}$) follows after an interchange of superscripts $l$ and $i$. We then obtain that

$$X_{(k)I(L)} = -X_{(k)L(I)} = x_{(k)}^{(+)l,(-)i} - x_{(k)}^{(+)i,(-)l}. \qquad (16)$$

Thus, the interbond charge transfer actually consists of contributions of separate pairs of formally-double bonds (I and L). (Besides, the contribution of the same (Ith) bond to $X_{(k)I}$ vanishes due to cancellation of increments of the right-hand side of Eq.(16)).

As opposed to the above-discussed extra population of the Ith double bond (that is divided equally between AOs $\chi_i^1$ and $\chi_{N+i}^2$ as Eq.(14) indicates), the last term of the right-hand side of Eq.(14) (i.e. $d_I$) offers contributions of opposite signs to occupation numbers of AOs $\chi_i^1$ and $\chi_{N+i}^2$, i.e. $d_I$ and $-d_I$, respectively. Thus, we have to do here with emergence of a dipole (polarization) inside the Ith double bond. Again, the increment $d_I$ coincides with the newly-emerging bond order between BOs $\phi_{(+)i}$ and $\phi_{(-)i}$ shown in Eq.(13).

It is seen, therefore, that the overall charge redistribution between AOs (atoms) due to our dual perturbation is representable as superposition of two additive components, viz. of the pairwise charge transfer between formally-double bonds (referred to below as the interbond one) and of polarization of the latter (i.e. emergence of intrabond dipoles). These components, in turn, originate from alterations in occupation numbers of BOs and from newly-formed bond orders between BOs inside individual double bonds, respectively.

To reveal the alterations in occupation numbers of AOs of polyenes due to introduction of heteroatom(s) themselves (we are interested in), we have now to separate the above-derived overall charge redistribution and/or its components into consequences of individual perturbations $\alpha$ and $\gamma$. To this end, let us consider the properties of the principal matrices $\boldsymbol{G}_{(k)}$, $\boldsymbol{D}_{(k)}^{(+)}$ and $\boldsymbol{D}_{(k)}^{(-)}$ of Eqs. (7) and (8) in a more detail.

*Analysis of interorbital interactions*

As already mentioned, elements of our principal matrices are chemically-meaningful. Let us dwell first just on this point. Let us start with matrices $\boldsymbol{G}_{(k)}$, $k=1,2\dots$ shown in Eq.(8). It is seen that the element $G_{(1)il}$ of the first order matrix $\boldsymbol{G}_{(1)}$ is proportional to the resonance parameter ($R_{il}$) between BOs $\phi_{(+)i}$ and $\phi_{(-)l}$ and inversely proportional to the relevant energy gap (equal to 2). Thus, $G_{(1)il}$ represents the direct (through-space) interaction of above-mentioned BOs. Accordingly, the diagonal element $G_{(1)ii}$ describes an analogous interaction between BOs $\phi_{(+)i}$ and $\phi_{(-)i}$ inside the Ith formally-double bond. Meanwhile, the element $G_{(2)il}$ of the second order matrix $\boldsymbol{G}_{(2)}$ is interpretable as the indirect (through-bond) interaction of the same BOs via a single mediator. Indeed, the second relation of Eq.(8) yields the following expression for this particular element

$$G_{(2)il} = -\frac{1}{2}[\sum_{(+)m} T_{im} G_{(1)ml} - \sum_{(-)m} G_{(1)im} Q_{ml}] \equiv \frac{1}{4}[\sum_{(+)m} T_{im} R_{ml} - \sum_{(-)m} R_{im} Q_{ml}], \qquad (17)$$

where sums over $(+)m$ and over $(-)m$ embrace all BBOs and all ABOs as previously. It is seen that both BBOs $\phi_{(+)m}$ and ABOs $\phi_{(-)m}$ are able to play the role of mediators of this interaction. In particular, a non-zero diagonal element $G_{(2)ii}$ (if any) implies an indirect intrabond interaction to arise between BOs $\phi_{(+)i}$ and $\phi_{(-)i}$, where orbitals of other double bonds are among potential mediators. Finally, the third order element $G_{(3)il}$ is interpretable analogously as the indirect interaction of BOs $\phi_{(+)i}$ and $\phi_{(-)l}$ by means of two mediators. Additivity of contributions of individual mediators to any of indirect interactions also deserves mention here.

Let us now turn to the series of matrices $\boldsymbol{D}_{(k)}^{(+)}$ and $\boldsymbol{D}_{(k)}^{(-)}$ defined by Eq.(7) and starting with $k=2$. As opposed to their former analogues $\boldsymbol{G}_{(k)}$, $k=2,3,..$, elements of the latter ($D_{(k)il}^{(+)}$ and $D_{(k)il}^{(-)}$) represent indirect interactions of orbitals inside the same subset, viz. between BBOs ($\phi_{(+)i}$ and $\phi_{(+)l}$) and between ABOs ($\phi_{(-)i}$ and $\phi_{(-)l}$), respectively. Accordingly, diagonal elements $D_{(k)ii}^{(+)}$ and $D_{(k)ii}^{(-)}$ coincide with indirect self-interactions of respective BOs. Thus, matrices $\boldsymbol{D}_{(k)}^{(+)}$ and $\boldsymbol{D}_{(k)}^{(-)}$ are correspondingly associated with subsets $\{\phi_{(+)}\}$ and $\{\phi_{(-)}\}$. It is also noteworthy that the above-specified indirect intrasubset interactions are mediated by a lower number of BOs as compared to intersubset ones underlying matrices $\boldsymbol{G}_{(k)}$, $k=2,3,..$ For example, BOs of the opposite subset only are able to play this role in the second order interactions $D_{(2)il}^{(+)}$ and $D_{(2)il}^{(-)}$, viz. ABOs ($\phi_{(-)m}$) and BBOs ($\phi_{(+)m}$), respectively.

Let us now employ the relations of Eqs.(7) and (8). The above interpretation of elements of matrices $\boldsymbol{G}_{(k)}$, $\boldsymbol{D}_{(k)}^{(+)}$ and $\boldsymbol{D}_{(k)}^{(-)}$ allows us then to conclude that an interaction of the second order (e.g. $G_{(2)il}$ or $D_{(2)il}^{(+)}$) originates from those of the first order (e.g. $G_{(1)im}$, see Eq.(17)). These newly-emerging second order interactions, in turn, potentially give birth to the third order ones (e.g. $G_{(3)il}$) and so forth. Again, the higher the parameter ($k$) is, the more orbitals of the molecule the interaction concerned embraces. Thus, a certain self-expansion of interactions of BOs is expected to take place due to perturbation.

More information about elements of our principal matrices $\boldsymbol{G}_{(k)}$, $\boldsymbol{D}_{(k)}^{(+)}$ and $\boldsymbol{D}_{(k)}^{(-)}$ follows after substituting Eq.(4) into Eqs.(7) and (8). Indeed, expressions for these matrices then consist of sums of members of different orders with respect to parameter $\alpha$ that may be analyzed separately. Let the relevant order parameter be denoted by the supersript ($s$). [Note that $s$ does not exceed $k$, i.e. $s=0,1,2,…k$]. The overall form of our matrices is then as follows

$$\boldsymbol{G}_{(k)} = \sum_{s=0}^{k}\boldsymbol{G}_{(k)}^{(s)}, \quad \boldsymbol{D}_{(k)}^{(+)} = \sum_{s=0}^{k}\boldsymbol{D}_{(k)}^{(s+)}, \quad \boldsymbol{D}_{(k)}^{(-)} = \sum_{s=0}^{k}\boldsymbol{D}_{(k)}^{(s-)}, \tag{18}$$

where increments of the same order ($s$) to matrices $\boldsymbol{D}_{(k)}^{(+)}$ and $\boldsymbol{D}_{(k)}^{(-)}$ are interrelated in addition, viz.

$$\boldsymbol{D}_{(k)}^{(0+)} = \boldsymbol{D}_{(k)}^{(0-)}, \qquad \boldsymbol{D}_{(k)}^{(1+)} = -\boldsymbol{D}_{(k)}^{(1-)}, \qquad \boldsymbol{D}_{(k)}^{(2+)} = \boldsymbol{D}_{(k)}^{(2-)}, etc \tag{19}$$

(The minus signs arise in these relations for terms of odd orders with respect to $\alpha$). It is also evident that the zero order ($\alpha$-independent) contributions of Eq.(18) (i.e. $\boldsymbol{G}_{(k)}^{(0)}$, $\boldsymbol{D}_{(k)}^{(0+)}$ and $\boldsymbol{D}_{(k)}^{(0-)}$) represent direct and indirect interactions of BOs of the parent polyene, whilst the remaining ($\alpha$-dependent) terms contain corrections to these interactions due to introduction of heteroatom(s). Furthermore, the above-exhibited additive form of matrices $\boldsymbol{G}_{(k)}$, $\boldsymbol{D}_{(k)}^{(+)}$ and $\boldsymbol{D}_{(k)}^{(-)}$ ensures also an analogous constitution of all related characteristics including total and partial transferred populations and intrabond dipoles as substitution of Eq.(18) into Eqs. (10), (12) and (13) shows. The same equally refers to the ultimate populations of AOs defined by Eqs.(14)-(16). This implies in summary that the $\alpha$-independent increments ($s=0$) and the $\alpha$-dependent corrections ($s=1,2,…k$) of our principal matrices correspondingly determine the initial charge distribution in the parent polyene and the subsequent redistribution due to introduction of heteroatom(s) and these two contributions may be studied separately. Let us consider the $\alpha$-independent terms at first.

Expressions for matrices $\boldsymbol{G}_{(k)}^{(0)}$, $k=1,2,3..$ resemble the definitions of their total analogues $\boldsymbol{G}_{(k)}$, $k=1,2,3…$ shown in Eqs.(8) and (17) except for zero order matrices with respect to parameter

$\alpha$ (viz. $T^{(0)}$, $Q^{(0)}$ and $R^{(0)}$ of Eq.(5)) standing instead of the total ones (i.e. $T$. $Q$ and $R$, respectively). This implies, for example, that $G_{(1)}^{(0)}$ is proportional to $R^{(0)}$ and, consequently, is a skew-symmetric matrix. Moreover, the same may be easily shown to refer to $G_{(2)}^{(0)}$, $G_{(3)}^{(0)}$, etc. [36]. As a result, diagonal elements of matrices $G_{(k)}^{(0)}$, $k=1,2,3..$ vanish (i.e. $G_{(k)ii}^{(0)} = 0$), whereas the off-diagonal ones are of opposite signs (i.e. $G_{(k)il}^{(0)} = -G_{(k)li}^{(0)}$). In respect of absolute values of the latter, however, matrices $G_{(1)}^{(0)}$, $G_{(2)}^{(0)}$ and $G_{(3)}^{(0)}$ differ one from another significantly. Thus, non-zero elements of the first order matrix $G_{(1)}^{(0)}$ ($G_{(1)il}^{(0)} \neq 0, i \neq l$) correspond only to BOs ($\phi_{(+)i}$ and $\phi_{(-)l}$) of first-neighbouring C=C bonds (I and L) and these are of uniform absolute values in addition coinciding with $\gamma/4$. Meanwhile, elements $G_{(2)il}^{(0)} \neq 0, i \neq l$ proved to be possible only for BOs $\phi_{(+)i}$ and $\phi_{(-)l}$ of second-neighbouring double bonds in polyenes [25, 26]. By contrast, $D_{(k)}^{(0+)}$ and $D_{(k)}^{(0-)}$ are symmetric matrices and contain non-zero diagonal elements $D_{(k)ii}^{(0+)}$ and $D_{(k)ll}^{(0-)}$. In particular, any second order element $D_{(2)ii}^{(0+)}$ is proportional to the total number of first neighbours of the C=C bond concerned ($n_I$), viz. $D_{(2)ii}^{(0+)} = \gamma^2 n_I / 16$, whereas the relevant third order corrections $D_{(3)ii}^{(0+)}$ take zero values [26].

Let us now turn to $\alpha$-dependent contributions of the same matrices $G_{(k)}$, $D_{(k)}^{(+)}$ and $D_{(k)}^{(-)}$ and start with the only correction of this type to direct interactions of BOs, i.e. $G_{(1)}^{(1)}$. As is seen from Eqs.(4) and (8), the latter is proportional to the matrix $R^{(1)}$ of Eq.(5). As a result, non-zero elements arise in the correction $G_{(1)}^{(1)}$ only in its diagonal positions referring to X=C and C=X bonds and these correspondingly coincide with $-\alpha/4$ and $\alpha/4$. Thus, introduction of a heteroatom into a certain (Ith) C=C bond of the parent polyene gives birth to a local direct interaction between the BBO and the ABO of this bond ($\phi_{(+)i}$ and $\phi_{(-)i}$). This result causes little surprise if we bear in mind that BOs of C=C bonds of the parent polyene $\{\phi\}$ defined by Eq.(2) are no longer eigenfunctions of the total Hamiltonian matrix block embracing an X=C( C=X) bond.

In contrast to the first order matrix $G_{(1)}$, its second order analogue $G_{(2)}$ contains two $\alpha$-dependent corrections $G_{(2)}^{(1)}$ and $G_{(2)}^{(2)}$ (see Eq.(18)) that are expressible via products of matrices of Eq.(5). In particular, the definition of the correction $G_{(2)}^{(2)}$ closely resembles that of the total matrix $G_{(2)}$ itself shown in Eq.(8) except for matrices $T^{(1)}$, $Q^{(1)}$ and $R^{(1)}$ standing for $T$, $Q$ and $R$, respectively. The diagonal constitution of the former along with coincidence between matrices $T^{(1)}$ and $Q^{(1)}$ seen from Eq.(5) then yields the equality $G_{(2)}^{(2)} = 0$. Meanwhile, the definition of the remaining correction $G_{(2)}^{(1)}$ contains all (i.e. six) matrices of Eq.(5). Moreover, $G_{(2)}^{(1)}$ is a symmetric matrix in contrast to the zero order increment $G_{(2)}^{(0)}$ of Eq.(18). An element of this matrix ($G_{(2)il}^{(1)}$) is expressible as follows

$$G_{(2)il}^{(1)} = \frac{1}{4}(T_{ii}^{(1)}R_{il}^{(0)} - R_{ii}^{(1)}Q_{il}^{(0)} + T_{il}^{(0)}R_{ll}^{(1)} - R_{il}^{(0)}Q_{ll}^{(1)}) \qquad (20)$$

and depends on elements of matrices of Eq.(5) referring to the Ith and the Lth double bonds only. At the same time, the correction $G_{(2)il}^{(1)}$ itself describes a certain additional indirect interaction between BOs of the same double bonds I and L ($\phi_{(+)i}$ and $\phi_{(-)l}$). Separate terms of the right-hand side of Eq.(20) correspondingly represent increments of four potential mediators of this interaction, viz. of BOs $\phi_{(+)i}$, $\phi_{(-)i}$, $\phi_{(+)l}$ and $\phi_{(-)l}$. Thus, the interacting orbitals $\phi_{(+)i}$ and $\phi_{(-)l}$ also are among potential mediators here, i.e. self-mediating effects of the latter are possible. The first order magnitude of the indirect interactions $G^{(1)}_{(2)il}$ with respect to parameter $\alpha$ also is easily seen from Eq.(20). Finally, a non-zero value condition may be formulated for the correction $G^{(1)}_{(2)il}$ on the basis of Eq.(20): First, presence of at least a single heteroatom within either of double bonds concerned (i.e. either within the Ith or within the Lth one) is among necessary conditions [Recall that significant diagonal elements of matrices $\boldsymbol{T}^{(1)}$, $\boldsymbol{Q}^{(1)}$ and $\boldsymbol{R}^{(1)}$ are entirely due to heteroatoms]. Second, a first neighbourhood of the Ith and Lth double bonds is imperative [As already mentioned, non-zero elements $T^{(0)}_{il}$, $Q^{(0)}_{il}$ and $R^{(0)}_{il}$ refer only to BOs of first-neighbouring C=C bonds in polyenes]. In summary, non-zero elements $G^{(1)}_{(2)il}$ are possible only between the BBO $\phi_{(+)i}$ of a heteroatom(s)-containing double bond (I) and the ABO $\phi_{(-)l}$ of its first neighbour (L) of any constitution (or vice versa), i.e. for two-membered conjugated fragments (I–L) containing at least a single heteroatom (X). [Note that indirect interactions between BOs of the same double bond (e.g. $G^{(1)}_{(2)ii}$) vanish because of zero diagonal elements of matrices $\boldsymbol{T}^{(0)}$, $\boldsymbol{Q}^{(0)}$ and $\boldsymbol{R}^{(0)}$]. In addition, the above-established local nature of elements $G^{(1)}_{(2)il}$ allows us to expect the relevant indirect interactions to be transferable for the same fragment I–L inside different compounds. [The above-formulated condition, however, is not sufficient to ensure a non-zero element $G^{(1)}_{(2)il}$. The point is that cancellation of terms of the right-hand side of Eq.(20) is possible as demonstrated below].

Let us now return to the first relation of Eq.(18) and recall that the zero order term $G^{(0)}_{(2)il}$ (the above-discussed correction $G^{(1)}_{(2)il}$ is added to) vanishes for first-neighbouring double bonds I and L in parent polyenes [25, 26]. Thus, a non-zero correction $G^{(1)}_{(2)il}$ (if any) implies emergence of a new indirect interaction between BOs of first-neighbouring double bonds I and L after introduction of heteroatom(s). It is also seen that local (direct) interactions of BOs inside a heteroatom-containing double bond (e. g. self-interactions of BOs inside the Ith bond ($T^{(1)}_{ii}$ and $Q^{(1)}_{ii}$) along with the relevant intrabond resonance parameter $R^{(1)}_{ii}$) potentially give birth to semi-local (indirect) interactions $G^{(1)}_{(2)il}$ embracing two-membered conjugated fragments I–L, the double bond concerned (i.e. the Ith one) participates in.

Let us now turn to $\alpha$-dependent increments of matrices $\boldsymbol{D}^{(+)}_{(k)}$ and $\boldsymbol{D}^{(-)}_{(k)}$ (see Eq.(18)). Due to simple relations of Eq.(19), we can confine ourselves to analysis of increments of matrices $\boldsymbol{D}^{(+)}_{(k)}$. Moreover, corrections $\boldsymbol{D}^{(1+)}_{(2)}$ and $\boldsymbol{D}^{(2+)}_{(2)}$ to the second order matrix $\boldsymbol{D}^{(0+)}_{(2)}$ prove to be sufficient for our purposes. The increment $\boldsymbol{D}^{(1+)}_{(2)}$ is the intrasubset analogue of the above-discussed correction $\boldsymbol{G}^{(1)}_{(2)}$. It is then no surprise that an element of this matrix ($D^{(1+)}_{(2)il}$) also depends on those of matrices of Eq.(5) pertinent to the Ith and Lth double bonds only, viz.

$$D^{(1+)}_{(2)il} = \frac{1}{4}(-R^{(1)}_{ii}R^{(0)}_{il} + R^{(0)}_{il}R^{(1)}_{ll}). \tag{21}$$

Analysis of Eq.(21) also closely resembles that of Eq.(20) except for a lower number of potential mediators of the underlying indirect interaction. Indeed, two BOs (viz. the ABOs $\phi_{(-)i}$ and $\phi_{(-)l}$) are now able to play this role instead of four BOs in the former case. [Note that the two terms of the right-hand side of Eq.(21) represent just the increments of ABOs $\phi_{(-)i}$ and $\phi_{(-)l}$, respectively]. In spite of the above-mentioned difference in the number of mediators, non-zero off-diagonal elements $D^{(1+)}_{(2)il}, i \neq l$ also refer to two-membered conjugated fragments (I–L) with at least a single heteroatom (X), whilst the relevant diagonal elements $D^{(1+)}_{(2)ii}$ take zero values. Finally, the remaining correction to the same second order matrix $D_{(2)}^{(0+)}$ (i.e. $D_{(2)}^{(2+)}$) is expressible as square of the above-considered matrix $G_{(1)}^{(1)}$. As with the latter, the matrix $D_{(2)}^{(2+)}$ then also contains non-zero elements of uniform absolute values in the diagonal positions associated with X=C(C=X) bonds, viz.

$$D_{(2)ii}^{(2+)} = (G_{(1)ii}^{(1)})^2 = \frac{\alpha^2}{16} > 0. \tag{22}$$

Let us now turn to $\alpha$-dependent increments of the third order matrix $G_{(3)}$, viz. $G_{(3)}^{(1)}$, $G_{(3)}^{(2)}$ and $G_{(3)}^{(3)}$ (see Eq.(18)). For reasons clarified below, we will confine ourselves here to analysis of diagonal elements $G_{(3)ii}^{(1)}$, $G_{(3)ii}^{(2)}$ and $G_{(3)ii}^{(3)}$. The increments $G_{(3)ii}^{(2)}$ vanish due to different positions of non-zero elements inside matrices contained within the respective expression (viz. $T^{(1)}$ and $G_{(2)}^{(1)}$, $G_{(1)}^{(1)}$ and $D_{(2)}^{(1+)}$, as well as $D_{(2)}^{(2+)}$ and $G_{(1)}^{(0)}$). Meanwhile, the analogous elements of the correction $G_{(3)}^{(3)}$ take the following simple form

$$G_{(3)ii}^{(3)} = -2D_{(2)ii}^{(2+)}G_{(1)ii}^{(1)} = -\frac{\alpha^2}{8}G_{(1)ii}^{(1)}, \tag{23}$$

where Eq.(22) also is used. It is seen that the element $G_{(3)ii}^{(3)}$ is proportional to the above-discussed direct intrabond interaction $G_{(1)ii}^{(1)}$. As a result, the former (i.e. $G_{(3)ii}^{(3)}$) also is of local and transferable nature and takes a non-zero value for any X=C(C=X) bond. Moreover, elements $G_{(3)ii}^{(3)}$ and $G_{(1)ii}^{(1)}$ are of opposite signs as Eq.(23) shows.

In contrast to Eq.(23), the expression for the remaining $\alpha$-dependent increment ($G_{(3)ii}^{(1)}$) formally contains a sum over BOs of all other double bonds (both BBOs $\phi_{(+)l}$ and ABOs $\phi_{(-)l}$), viz.

$$G_{(3)ii}^{(1)} = \sum_l (2D_{(2)il}^{(1+)}G_{(1)il}^{(0)} - T_{il}^{(0)}G_{(2)il}^{(1)}) - \frac{\gamma^2 n_I}{8}G_{(1)ii}^{(1)}, \tag{24}$$

where the above-discussed proportionality of elements $D_{(2)ii}^{(0+)}$ to total numbers of first neighbours of double bonds concerned ($n_I$) is additionally invoked. Actually, however, the sum of Eq.(24) embraces BOs only of first neighbours of the Ith double bond [As discussed already, non-zero elements of matrices $G_{(1)}^{(0)}$, $T^{(0)}$, $G_{(2)}^{(1)}$ and $D_{(2)}^{(1+)}$ belong to first-neighbouring pairs of bonds]. In other words, the element $G_{(3)ii}^{(1)}$ consists of contributions of two-membered conjugated fragments I–L, the Ith double bond participates in.

On the whole, $\alpha$-dependent corrections of two types reveal themselves in the above-analyzed matrix elements, viz. i) local corrections associated with individual X=C(C=X) bonds (such as $G_{(1)ii}^{(1)}$, as well as $D_{(2)ii}^{(2+)}$ and $G_{(3)ii}^{(3)}$ of Eqs.(22) and (23), respectively) and ii) semi-local corrections embracing two-membered conjugated fragments I–L (viz. $G_{(2)il}^{(1)}$, $D_{(2)il}^{(1+)}$ and $G_{(3)ii}^{(1)}$ of Eqs.(20), (21) and (24), respectively). Accordingly, elements of matrices of higher orders ($k$=4,5,..) may be easily shown to contain $\alpha$-dependent corrections extending over larger fragments of polyenes with heteroatoms.

Let us now turn to analysis of increments to both partial transferred populations and intrabond dipoles originating from each of the above-revealed corrections to interorbital interactions separately.

*Local and semi-local contributions to the charge redistribution*

Let us start with implications of the above-discussed $\alpha$-independent matrices $\boldsymbol{G}^{(0)}_{(k)}$, $\boldsymbol{D}^{(0+)}_{(k)}$ and $\boldsymbol{D}^{(0-)}_{(k)}$ for charge distribution in the parent polyenes: First, dipoles of Eq.(13) vanish for all C=C bonds in these hydrocarbons in accordance with the expectation, i.e. $d^{(0)}_{(k)I} = 0$ for $k=1,2,3\ldots$ and thereby $d^{(0)}_I = 0$. Second, the relevant partial transferred populations meet the following symmetry relations

$$x^{(0),(+)l,(-)i}_{(k)} = x^{(0),(+)i,(-)l}_{(k)}, \qquad k = 2,3\ldots \tag{25}$$

and, consequently, both $X^{(0)}_{(k)I(L)}$ and $X^{(0)}_{(k)L(I)}$ of Eq.(16) take zero values. Hence, no interbond charge transfer actually takes place in the parent polyenes and this result also causes little surprise. At the same time, the partial transferred populations of Eq.(25) themselves generally do not vanish. In particular, the first substantial members of the respective power series (i.e. $x^{(0),(+)l,(-)i}_{(2)}$ and $x^{(0),(+)i,(-)l}_{(2)}$) take non-zero values for first-neighbouring C=C bonds I and L. As a result, an exchange of population between these bonds may be concluded to underly the zero ultimate interbond charge transfer. Meanwhile, the subsequent corrections $x^{(0),(+)l,(-)i}_{(3)}$ and $x^{(0),(+)i,(-)l}_{(3)}$ vanish because no bonds I and L are present in polyenes that are both first and second neighbours simultaneously [26].

Let us now turn to consequences of $\alpha$-dependent contributions of matrices $\boldsymbol{G}_{(k)}$, $\boldsymbol{D}^{(+)}_{(k)}$ and $\boldsymbol{D}^{(-)}_{(k)}$.

Let us start with partial transferred populations defined by Eq.(12). The population of the second order ($x^{(+)i,(-)l}_{(2)}$) formally contains two $\alpha$-dependent increments, viz. $x^{(1),(+)i,(-)l}_{(2)}$ and $x^{(2),(+)i,(-)l}_{(2)}$, the first one actualy vanishing due to different positions of non-zero elements inside matrices $\boldsymbol{G}^{(0)}_{(1)}$ and $\boldsymbol{G}^{(1)}_{(1)}$. Meanwhile, the second increment takes a non-zero value inside any X=C(C=X) bond (i.e. for $l=i$), viz.

$$x^{(2),(+)i,(-)i}_{(2)loc} = 2(G^{(1)}_{(1)ii})^2 = \frac{\alpha^2}{8} > 0. \tag{26}$$

The correction of Eq.(26) represents the population transferred from the BBO $\phi_{(+)i}$ to the ABO $\phi_{(-)i}$ due to introduction of a single heteroatom into the Ith C=C bond of the parent polyene. The sub- or superscript *loc* here and below indicates just the local (i.e. intrabond) nature of the relevant increment. It also deserves recalling here that any charge transfer of intrabond type does not contribute to alterations in populations of AOs (see Eqs.(14) and (16) and the discussion nearby).

To reveal the effect of heteroatom(s) upon the interbond charge transfer, let us now look for $\alpha$-dependent increments within the partial transferred populations of the third order, viz. $x^{(+)i,(-)l}_{(3)}$ of Eq.(12). The contributions $x^{(2),(+)i,(-)l}_{(3)}$ and $x^{(3),(+)i,(-)l}_{(3)}$ also vanish here for analogous reasons, whereas the remaining one (i.e. $x^{(1),(+)i,(-)l}_{(3)}$) potentially takes a non-zero value and meets the following anti-symmetry relation in addition

$$x^{(1),(+)i,(-)l}_{(3)} = -x^{(1),(+)l,(-)i}_{(3)}, \tag{27}$$

where

$$x^{(1),(+)i,(-)l}_{(3)} = 4G^{(0)}_{(1)il}G^{(1)}_{(2)il}, \qquad x^{(1),(+)l,(-)i}_{(3)} = 4G^{(0)}_{(1)li}G^{(1)}_{(2)li}. \tag{28}$$

[Recall that $G^{(0)}_{(1)il} = -G^{(0)}_{(1)li}$ and $G^{(1)}_{(2)il} = G^{(1)}_{(2)li}$]. The opposite nature of the above relation vs. that referring to the parent hydrocarbon (see Eq.(25)) deserves emphasizing here. As is seen from Eq.(27), the additional population transferred between orbitals $\phi_{(+)i}$ and $\phi_{(-)l}$ and its counterpart referring to the remaining BOs $\phi_{(+)l}$ and $\phi_{(-)i}$ are now of opposite signs and of uniform absolute values. For example, if the BBO $\phi_{(+)i}$ acquires a certain population from the ABO $\phi_{(-)l}$ (i.e.

$x_{(3)}^{(1),(+)i,(-)l} <0$), an analogous population is transferred from the BBO $\phi_{(+)l}$ to the ABO $\phi_{(-)i}$ (i.e. $x_{(3)}^{(1),(+)l,(-)i} >0$) or vicee versa. In other words, population is predicted to be unambiguosly transferred either from the Ith bond to the Lth one or vice versa due to introduction of heteroatom(s). The ultimate direction of this charge transfer depends on signs of elements $G_{(1)il}^{(0)}$ and $G_{(2)il}^{(1)}$: Given that these signs are uniform (opposite), the Ith double bond looses (acquires) an additional population and plays the role of the electron-donating (accepting) bond with respect to the Lth one. Obviously, no charge transfer is expected for pairs of double bonds characterized by zero corrections $G_{(2)il}^{(1)}$. If we recall finally that elements both $G_{(2)il}^{(1)}$ and $G_{(1)il}^{(0)}$ take non-zero values for pairs of first-neighbouring double bonds I and L only and thereby for two-membered conjugated fragments I–L, significant partial transferred populations $x_{(3)}^{(1),(+)i,(-)l}$ also are associated exclusively with individual fragments I–L and are among transferable characteristics of the latter. Finally, employment of Eq. (27) within Eq.(16) shows that the respective population acquired (donated) by the Ith bond from (to) the Lth one ($X_{(3)I(L)}^{(1)}$) coincides with the two-fold correction under discussion, i.e. with $2 x_{(3)}^{(1),(+)l,(-)i}$.

Let us now dwell on $\alpha$-dependent contributions to intrabond dipoles of Eq.(13). The above-discussed non-zero elements $G_{(1)ii}^{(1)}$ evidently give birth to local and transferable dipoles expressible as follows

$$d_{(1)I}^{(1)loc} = -2G_{(1)ii}^{(1)} = R_{ii}^{(1)} = \pm \frac{\alpha}{2}, \qquad (29)$$

where the upper and lower signs correspondingly refer to X=C and C=X bonds. As is seen from Eqs.(14) and (29), these dipoles contribute to growth and reduction of populations of AOs of X and C atoms, respectively, of the Ith double bond in accordance with the expectation for a more electronegative heteroatom X (i.e. for $\alpha >0$). Meanwhile, the subsequent two increments of Eq.(13) (viz. $d_{(2)I}^{(1)}$ and $d_{(2)I}^{(2)}$) vanish due to zero elements $G_{(2)ii}^{(1)}$ and $G_{(2)ii}^{(2)}$. Thus, we can turn immediately to consequences of diagonal elements of the third order matrix $\mathbf{G}_{(3)}$ to bond dipoles, i.e. to increments $d_{(3)I}$ of Eq.(13).

As discussed already, the element $G_{(3)ii}$ consists of two significant $\alpha$-dependent increments shown in Eq.(23) and (24). The first one (i.e. $G_{(3)ii}^{(3)}$ of Eq.(23)) gives birth to an additional local and transferable contribution to the total dipole of any X=C(C=X) bond, viz.

$$d_{(3)I}^{(3)loc} = -2G_{(3)ii}^{(3)} = \frac{\alpha^2}{4} G_{(1)ii}^{(1)} = -\frac{\alpha^2}{8} R_{ii}^{(1)} = \mp \frac{\alpha^3}{16} \qquad (30)$$

that takes an opposite direction vs. that of the former increment $d_{(1)I}^{(1)loc}$ of Eq.(29). [Note that the meaning of signs of the right-hand side of Eq.(30) coincides with that of Eq.(29)]. As a result, the sum of increments of Eqs.(29) and (30) also is of local nature and coincides with the dipole of an isolated X=C(C=X) bond to within the third order terms with respect $\alpha$. That is why the above-specified sum may be referred to as the primary dipole of an X=C(C=X) bond.

By contrast, the remaining increment under our interest (i.e. $G_{(3)ii}^{(1)}$ of Eq.(24)) is expected to give birth to non-local (i.e. environment-dependent) dipoles of formally-double bonds that may be alternatively referred to as secondary or induced ones. To make sure that this is the case, let us recall the constitution of the element $G_{(3)ii}^{(1)}$ as a sum of contributions of individual first neighbours (L) of the Ith bond seen from Eq.(24). The same then consequently refers to the related dipole $d_{(3)I}^{(1)}$. Let the partial contribution of the Lth first neighbour to this dipole be denoted by $d_{(3)I(L)}^{(1)}$. From Eqs. (13) and (24), we the obtain

$$d^{(1)}_{(3)I(L)} = 2T^{(0)}_{il}G^{(1)}_{(2)il} - 4D^{(1+)}_{(2)il}G^{(0)}_{(1)il} + \frac{\gamma^2}{4}G^{(1)}_{(1)ii}. \tag{31}$$

Accordingly, the expression for the contribution of the Ith first neighbour to the analogous dipole of the Lth double bond ($d^{(1)}_{(3)L(I)}$) follows from Eq.(31) after an interchange of subscripts $i$ and $l$, as well as $I$ and $L$. We then obtain that

$$d^{(1)}_{(3)L(I)} = 2T^{(0)}_{li}G^{(1)}_{(2)li} - 4D^{(1+)}_{(2)li}G^{(0)}_{(1)li} + \frac{\gamma^2}{4}G^{(1)}_{(1)ll}. \tag{32}$$

It is seen that matrix elements contained within Eqs.(31) and (32) embrace BOs of the Ith and Lth double bonds only and are independent of the remaining part of the molecule. Moreover, the first two members of the right-hand sides of these expressions are interrelated as discussed below, whereas the last ones depend on characteristics of each double bond (I and/or L) separately. Let us now consider these points in a more detail.

The first members of expressions for $d^{(1)}_{(3)I(L)}$ and $d^{(1)}_{(3)L(I)}$ coincide one with another due to the above-established symmetric nature of matrices $T^{(0)}$ and $G^{(1)}_{(2)}$. Hence, uniform and parallel dipoles arise due to these terms inside both formally-double bonds I and L simultaneously. Emergence of these dipoles, in turn, implies a significant total polarization of the whole fragment I–L that may be called the longitudinal one. Meanwhile, the second members of Eqs.(31) and (32) are of the same absolute value but of opposite signs and consequently yield similar but anti-parallel dipoles inside the same double bonds [Recall that $G^{(0)}_{(1)}$ is a skew-symmetric matrix]. This evidently implies a zero consequent polarity of the fragment I–L. The relevant polarization is then referred to below as the internal one.

Let us introduce the following designations for the longitudinal and the internal polarizations of the fragment I–l, respectively, viz.

$$p^{(1)long}_{(3)I-L} = 2T^{(0)}_{il}G^{(1)}_{(2)il} = 2T^{(0)}_{li}G^{(1)}_{(2)li}; \qquad p^{(1)int}_{(3)I-L} = -4D^{(1+)}_{(2)il}G^{(0)}_{(1)il} = 4D^{(1+)}_{(2)li}G^{(0)}_{(1)li}. \tag{33}$$

As is seen from Eq.(33), non-zero elements $G^{(1)}_{(2)il}$ and $D^{(1+)}_{(2)il}$ correspondingly are the necessary conditions for the above-specified two types of polarization to arise. Since these elements are characteristics of the whole fragment I–L, the same equally refers also to consequent polarizations. Moreover, a certain relation may be anticipated between the longitudinal polarization ($p^{(1)long}_{(3)I-L}$) and the intrafragmental charge transfer described by Eqs.(27) and (28).

Let us now dwell on the last increments of the right-hand sides of Eqs. (31) and (32) that depend on respective local direct intrabond interactions ($G^{(1)}_{(1)ii}$ and $G^{(1)}_{(1)ll}$) and thereby are independent one from another in contrast to the above-considered members of the same relations. This implies the increments concerned to give birth to certain additional dipoles of our double bonds I and L, provided that the latter are X=C or C=X ones. Let it be the case with the Ith bond. It is then seen that the direction of the respective new dipole coincides with that of $d^{(3)loc}_{(3)I}$ of Eq.(30) and, consequently, an extra reduction of the relevant primary dipole is anticipated. As opposed to $d^{(3)loc}_{(3)I}$, however, the increment under discussion depends also on the interbond resonance parameter $\gamma$ (see Eq. (31)). This fact gives us a hint that the new dipole actually is of non-local nature and arises only under presence of other bonds. Reduction of the primary dipole of any heteropolar bond after incorporating the latter into a certain molecule (the so-called depolarization) has been studied previously in a detail [38, 39] and proved to be a rather universal phenomenon. In this connection, the new dipole of the Ith bond (i.e. that originating from the last member of Eq.(31)) is accordingly interpretable as the contribution of the Lth first neighbour (and/or of the fragment I–L) to the total depolarization dipole of the Ith X=C or C=X bond. Let this contribution be denoted by $d_{(3)I(L)}^{(1)dep}$. We then obtain

$$d_{(3)I(L)}^{(1)dep} = \frac{\gamma^2}{4} G_{(1)ii}^{(1)} = -\frac{\gamma^2}{8} R_{ii}^{(1)} = \mp \frac{\alpha\gamma^2}{16}. \tag{34}$$

As is seen from the above relation, the contribution $d_{(3)I(L)}^{(1)dep}$ actually is independent of the nature of the relevant neighbour L itself. This implies that contributions of all first neighbours of the Ith X=C or C=X bond to the total depolarization dipole of the latter ($d_{(3)I}^{(1)dep}$) are uniform and, consequently, $d_{(3)I}^{(1)dep}$ is proportional to the number of these neighbours ($n_I$) (see also Eq.(24)).

Let us now return to Eqs.(31) and (32) and take into consideration the above-suggested interpretations of separate members of these relations. It is then seen that $d_{(3)I(L)}^{(1)}$ of Eq.(31) represents the contribution of the Lth first neighbour (and/or of the fragment I–L) to the total secondary (induced) dipole of the Ith double bond ($d_{(3)I}^{(1)}$). Accordingly, $d_{(3)L(I)}^{(1)}$ of Eq.(32) coincides with the contribution of the Ith first neighbour (and thereby of the same fragment I–L) to the analogous dipole of the Lth double bond ($d_{(3)L}^{(1)}$). It is also evident that contributions of the given fragment I–L to secondary dipoles of the Ith and Lth double bonds generally differ one from another as illustrated in the next Subsection. Meanwhile, the main conclusion of this Subsection (embracing both the partial transferred populations and intrabond dipoles) is that the overall charge redistribution in polyenes due to introduction of heteroatom(s) consists of local redistributions inside individual X=C(C=X) bonds and of semi-local ones associated with separate two-membered conjugated fragments I–L. The environment-dependent depolarization dipoles of X=C(C=X) bonds take an intermediate place in this classification.

Now, we are about to focus our attention on semi-local increments of Eqs.(28) and (33) that are determined by interbond interactions $G_{(2)il}^{(1)}$ and $D_{(2)il}^{(1+)}$ pertinent to the fragment I–L and thereby potentially depending on the actual constitution of the latter. The next Subsection addresses just this dependence and its consequences upon absolute values and signs of contributions of different fragments I–L to the charge redistribution.

*Comparison of contributions of separate conjugated fragments*

Nine different two-membered conjugated fragments (I–L) [40] are possible in polyenes with heteroatom(s), viz.

X = C – C = C   (I)     C = X – C = C   (II)    X = C – X = C   (III)
X = C – C = X   (IV)    C = X – X = C   (V)     X = X – C = C   (VI)         (35)
X = X – X = C   (VII)   X = X – C = X   (VIII)  X = X – X = X   (IX)

Obviously, fragments I-IX coincide with derivatives of butadiene containing different numbers of heteroatoms (X) in various positions. Let the four carbon atoms of the parent butadiene be enumerated as follows

$$C_i^1 = C_{N+i}^2 - C_l^1 = C_{N+l}^2, \tag{36}$$

where the 2p$_z$ AOs of carbon atoms under numbers $i$, $l$ and $N+i$, $N+l$ are supposed to belong to subsets $\{\chi^1\}$ and $\{\chi^2\}$, respectively, as indicated by superscripts. The same numbering of AOs and/or of atoms will be preserved also when passing to fragments I-IX of Eq.(35). As a result, elements of $\alpha$-independent matrices of Eqs.(4), (5) and (8) are as follows

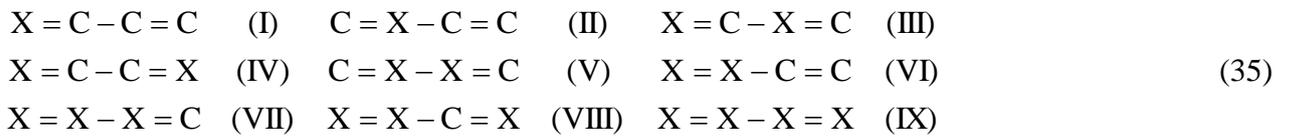

$$T_{il}^{(0)} = -Q_{il}^{(0)} = R_{il}^{(0)} = -R_{li}^{(0)} = \frac{\gamma}{2}, \qquad G_{(1)il}^{(0)} = -G_{(1)li}^{(0)} = -\frac{\gamma}{4} \tag{37}$$

for all fragments concerned. Meanwhile, elements of matrices $\boldsymbol{T}^{(1)}$, $\boldsymbol{Q}^{(1)}$ and $\boldsymbol{R}^{(1)}$ depend on the nature of double bonds I and L inside the given frgment I–L (see the discussion shortly after Eq.(5)) and, consequently, these generally take distinct values for separate fragments. [Besides, the collection of these elements for fragments I-IX may be found in Table 2 of Ref.[40]]. As is seen from Eqs.(20) and (21), knowledge of the above-enumerated elements is sufficient for analysis of $\alpha$-dependent interactions $G^{(1)}_{(2)il}$ and $D^{(1+)}_{(2)il}$ for fragments I-IX.

Let us start with mono-substituted fragments I and II, wherein the Ith C=C bond of the parent butadiene is replaced by an X=C bond and by a C=X one, respectively. In other terms, the only heteroatom X correspondingly takes the $i$th and the $(N+i)$th position inside the parent fragment of Eq.(36). The principal difference between fragments I and II then consequently consists in opposite signs of the intrabond resonance parameters ($R^{(1)}_{ii}$) between BOs of the Ith double bond ($\phi_{(+)i}$ and $\phi_{(-)i}$) that coincide with $\alpha/2$ and $-\alpha/2$, respectively. This seemingly ordinary circumstance, however, yields dramatic consequences upon $\alpha$-dependent indirect interactions of BOs.

To discuss this point in a more detail, let us turn to the expression of Eq.(20) for the correction $G^{(1)}_{(2)il}$. Employment of the above-overviewed elements of matrices of Eqs.(4) and (5) shows that the last two contributions of the right-hand side of Eq.(20) vanish for both fragments concerned, whereas the remaining members are of non-zero and uniform absolute values. More importantly, the latter are of the same and of opposite signs for fragments I and II, respectively. Thus, the mediating effects of BOs $\phi_{(+)i}$ and $\phi_{(-)i}$ correspondingly are added together and cancel out one another in the cases I and II. As a result, a non-zero correction $G^{(1)}_{(2)il}$ arises for the first fragment (I) only as shown in Table 1. At the same time, the relevant indirect interactions between BBOs (i.e. the corrections $D^{(1+)}_{(2)il}$) are of uniform absolute values and of opposite signs for fragments I and II (Table 1) as turning to Eq.(21) shows [Note that the ABO $\phi_{(-)i}$ is the only mediator in this case]. Besides, an analogous state of things (i.e. addition and cancellation of separate increments to the indirect interaction between a BBO and an ABO) has been observed also when comparing linear and cross-conjugated isomers of polyenes [25].

Table 1

Contributions of two-membered conjugated fragments I–L (I-IX) to total secondary (induced) dipoles of the Ith and Lth double bonds ($d^{(1)}_{(3)I(L)}$ and $d^{(1)}_{(3)L(I)}$, respectively) along with their separate components, viz. the longitudinal polarization ($p^{(1)long}_{(3)I-L}$), the internal counterpart of the latter ($p^{(1)int}_{(3)I-L}$) and the relevant depolarization increments ($d^{(1)dep}_{(3)I(L)}$ and $d^{(1)dep}_{(3)L(I)}$). Additional indirect interactions of BOs ($G^{(1)}_{(2)il}$ and $D^{(1+)}_{(2)il}$) also are shown

| Nr. | $d^{(1)}_{(3)I(L)}$ | $d^{(1)}_{(3)L(I)}$ | $p^{(1)long}_{(3)I-L}$ * | $p^{(1)int}_{(3)I-L}$ | $d^{(1)dep}_{(3)I(L)}$ | $d^{(1)dep}_{(3)L(I)}$ | $G^{(1)}_{(2)il}$ | $D^{(1+)}_{(2)il}$ |
|---|---|---|---|---|---|---|---|---|
| I | 0 | $\dfrac{3\alpha\gamma^2}{16}$ | $\dfrac{\alpha\gamma^2}{8}$ | $-\dfrac{\alpha\gamma^2}{16}$ | $-\dfrac{\alpha\gamma^2}{16}$ | 0 | $\dfrac{\alpha\gamma}{8}$ | $-\dfrac{\alpha\gamma}{16}$ |
| II | $\dfrac{2\alpha\gamma^2}{16}$ | $-\dfrac{\alpha\gamma^2}{16}$ | 0 | $\dfrac{\alpha\gamma^2}{16}$ | $\dfrac{\alpha\gamma^2}{16}$ | 0 | 0 | $\dfrac{\alpha\gamma}{16}$ |
| III | $\dfrac{\alpha\gamma^2}{16}$ | $\dfrac{\alpha\gamma^2}{16}$ | $\dfrac{\alpha\gamma^2}{8}$ | 0 | $-\dfrac{\alpha\gamma^2}{16}$ | $-\dfrac{\alpha\gamma^2}{16}$ | $\dfrac{\alpha\gamma}{8}$ | 0 |
| IV | $-\dfrac{3\alpha\gamma^2}{16}$ | $\dfrac{3\alpha\gamma^2}{16}$ | 0 | $-\dfrac{\alpha\gamma^2}{8}$ | $-\dfrac{\alpha\gamma^2}{16}$ | $\dfrac{\alpha\gamma^2}{16}$ | 0 | $-\dfrac{\alpha\gamma}{8}$ |
| V | $\dfrac{3\alpha\gamma^2}{16}$ | $-\dfrac{3\alpha\gamma^2}{16}$ | 0 | $\dfrac{\alpha\gamma^2}{8}$ | $\dfrac{\alpha\gamma^2}{16}$ | $-\dfrac{\alpha\gamma^2}{16}$ | 0 | $\dfrac{\alpha\gamma}{8}$ |
| VI | $\dfrac{2\alpha\gamma^2}{16}$ | $\dfrac{2\alpha\gamma^2}{16}$ | $\dfrac{\alpha\gamma^2}{8}$ | 0 | 0 | 0 | $\dfrac{\alpha\gamma}{8}$ | 0 |
| VII | $\dfrac{3\alpha\gamma^2}{16}$ | 0 | $\dfrac{\alpha\gamma^2}{8}$ | $\dfrac{\alpha\gamma^2}{16}$ | 0 | $-\dfrac{\alpha\gamma^2}{16}$ | $\dfrac{\alpha\gamma}{8}$ | $\dfrac{\alpha\gamma}{16}$ |
| VIII | $-\dfrac{\alpha\gamma^2}{16}$ | $\dfrac{2\alpha\gamma^2}{16}$ | 0 | $-\dfrac{\alpha\gamma^2}{16}$ | 0 | $\dfrac{\alpha\gamma^2}{16}$ | 0 | $-\dfrac{\alpha\gamma}{16}$ |
| IX | 0 | 0 | 0 | 0 | 0 | 0 | 0 | 0 |

*The longitudinal polarization ($p_{(3)I-L}^{(1)long}$) coincides with the partial population transferred between the Ith and Lth double bonds inside the given fragment I–L ($x_{(3)}^{(1),(+)l,(-)i}$).

To consider the di-substituted fragments, let us start with the fragment III consisting of two X=C bonds. The relevant $\alpha$-dependent matrices $\boldsymbol{T}^{(1)}$ ($\boldsymbol{Q}^{(1)}$) and $\boldsymbol{R}^{(1)}$ then contain non-zero elements in both the $i$th and $l$th positions. As a result, four non-zero increments of uniform absolute values arise in the expressions of Eq.(20) for $G_{(2)il}^{(1)}$, i.e. four BOs (viz. $\phi_{(+)i}$, $\phi_{(-)i}$, $\phi_{(+)l}$ and $\phi_{(-)l}$) are able to mediate the indirect interaction between BOs $\phi_{(+)i}$ and $\phi_{(-)l}$. Analysis of signs of these increments, however, shows that mediating effects of BOs $\phi_{(+)i}$ and $\phi_{(-)i}$ are added together (as it was the case with the fragment I), whereas the contributions of BOs $\phi_{(+)l}$ and $\phi_{(-)l}$ cancel out one another (in analogy with the fragment II). The ultimate correction $G_{(2)il}^{(1)}$(III) then consequently coincides with that of the fragment I (Table 1). At the same time, the $\alpha$-dependent correction to the indirect interaction between BBOs (i.e. $D_{(2)il}^{(1+)}$ defined by Eq.(21)) vanishes for the fragment III due to cancellation of increments of two mediators ($\phi_{(-)i}$ and $\phi_{(-)l}$).

As with the above-discussed fragment III, all the four BOs are among potential mediators of indirect interactions $G_{(2)il}^{(1)}$ referring to the subsequent di-substituted fragments IV and V. Moreover, there are two positive and two negative contributions in the relevant expressions of Eq.(20), e.g. positive increments of BOs $\phi_{(+)i}$ and $\phi_{(-)i}$ to $G_{(2)il}^{(1)}$(IV) go together with negative ones of BOs $\phi_{(+)l}$ and $\phi_{(-)l}$. Consequently, the corrections $G_{(2)il}^{(1)}$ vanish for both fragments concerned. By contrast, contributions of both mediators (i.e. of ABOs $\phi_{(-)i}$ and $\phi_{(-)l}$) are added together in the expressions of Eq.(21) for corrections $D_{(2)il}^{(1+)}$(IV) and $D_{(2)il}^{(1+)}$(V) so that these consequently take double higher absolute values as compared to $D_{(2)il}^{(1+)}$(I) and $D_{(2)il}^{(1+)}$(II). At the same time, the above-mentioned corrections are of opposite signs in analogy with the mono-substituted fragments I and II.

As opposed to the above-considered di-substituted fragments III-V, the last one (VI) contains an X=X bond represented by a zero intrabond resonance parameter $R_{ii}^{(1)}$. Meanwhile, elements of the remaining $\alpha$-dependent matrices of Eq.(5) (i.e. of $T_{ii}^{(1)}$ and $Q_{ii}^{(1)}$) take double values ($\alpha$) vs. those associated with C=X bonds. Turning to Eq.(20) then shows th BBO $\phi_{(+)i}$ actually is the only mediator of the indirect interaction $G_{(2)il}^{(1)}$(VI) and the latter ultimately coincides with the relevant corrections for fragments I and III. Again, employment of the equality $R_{ii}^{(1)}=0$ yields a vanishing correction $D_{(2)il}^{(1+)}$(VI).

Fragments with three heteroatoms (VII and VIII) offer no new aspects to the present discussion due to their similarity to mono-substituted ones (i.e. to I and II, respectively). This result causes little surprise if we observe that the only carbon atom of fragments VII and VIII plays the role of the heteroatom here. Accordingly, the last fragment (IX) is an analogue of the parent butadiene.

It is seen, therefore, that the indirect interactions between the BBO $\phi_{(+)i}$ and the ABO $\phi_{(-)l}$ ($G_{(2)il}^{(1)}$) meet a simple selection rule for above-considered fragments I-IX, viz. these coincide with a constant ($\alpha\gamma/8$) for fragments I, III, VI and VII and take a zero value for the remaining ones (II, IV, V, VIII and IX). These groups of fragments are referred to below as the first and the second one. Meanwhile, the relevant indirect interactions between BBOs ($D_{(2)il}^{(1+)}$) exhibit more diversity in respect of both absolute values and signs. In particular, these interactions are of uniform absolute values and of opposite signs for pairs of fragments I and II, IV and V, as well as VII and VIII and vanish for the remaining ones (III, VI and IX).

Let us now turn to contributions of individual fragments to the charge redistribution in polyenes defined by Eqs.(28) and (33). An additional employment of Eq.(37) then shows that both the partial population transferred inside a certain fragment I–L ($x_{(3)}^{(1),(+)l,(-)i}$) and the respective longitudinal polarization ($p_{(3)I-L}^{(1)long}$) are expressible as $\gamma G_{(2)il}^{(1)}$ and thereby coincide one with another. Hence, the above-specified increments to charge redistribution also meet an analogous simple selection rule, viz. these equal to a positive constant ($\alpha\gamma^2/8$) for a more electronegative heteroatom ($\alpha>0$) and to zero for fragments of the first and of the second group, repectively (Table 1). In other terms, fragments I, III, VI and VII are characterized by a significant charge transfer from the right double bond (L) to the left one (I) for $\alpha>0$ and their longitudinal polarization is orientated similarly, e.g.

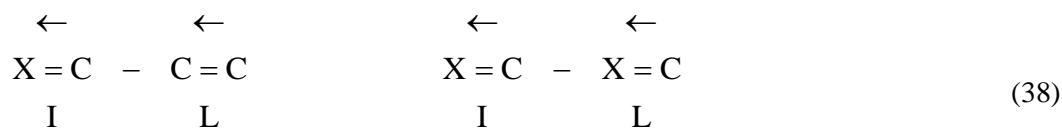

$$\begin{array}{cccc} \leftarrow & \leftarrow & \leftarrow & \leftarrow \\ X=C & - & C=C & \quad X=C & - & X=C \\ I & & L & \quad I & & L \end{array} \qquad (38)$$

It is also seen that the positive longitudinal polarirization exhibited in Eq.(38) contributes to growth in total dipoles of X=C bonds. Meanwhile, neither an intrafragmental charge transfer nor a longitudinal polarization are peculiar to fragments of the second group (II, IV, V, VIII and IX).

By contrast, the internal polarizations of fragments I-IX ($p_{(3)I-L}^{(1)int}$) are expressible as $\gamma D_{(2)il}^{(1+)}$ (see Eqs. (33) and (37)). As a result, $p_{(3)I-L}^{(1)int}$ take positive and negative values for fragments II, V, VII and I, IV, VIII, respectively, and vanish for the remaining ones (III, VI, IX) (Table 1). The consequent directions of the bond dipoles are exemplified below

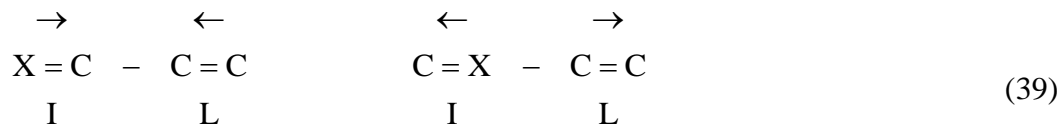

$$X = C \; - \; C = C \qquad C = X \; - \; C = C \qquad (39)$$
$$\;\; I \qquad\quad L \qquad\qquad\;\; I \qquad\quad L$$

It is also noteworthy that the actual signs of $p^{(1)int}_{(3)I-L}$ reflect a trend towards reduction of primary dipoles of X=C(C=X) bonds in analogy with the relevant increments to depolarization dipoles ($d^{(1)dep}_{(3)I(L)}$ and $d^{(1)dep}_{(3)L(I)}$).

Total contributions of fragments I-IX to secondary (induced) dipoles of the Ith and Lth double bonds ($d^{(1)}_{(3)I(L)}$ and $d^{(1)}_{(3)L(I)}$) easily follow from Eqs.(31) and (32), respectively, and are shown in the left segment of Table 1. For fragments of the first group, the longitudinal polarization predominates over the remaining members of the above-mentioned relations. As a result, the relevant total contributions $d^{(1)}_{(3)I(L)}$ and $d^{(1)}_{(3)L(I)}$ take positive values and point to shifts of population from the right to the left inside both the Ith and Lth bonds of fragments concerned as exhibited in Eq.(38) [$d^{(1)}_{(3)I(L)}$(I) and $d^{(1)}_{(3)L(I)}$(VII) make an exception and coincide with zero]. Meanwhile, the above-concluded reduction of primary dipoles of X=C and C=X bonds proves to be the ultimate result for fragments of the second group due to absence of the longitudinal polarization.

Emergence of secondary (induced) dipoles of initially-homopolar (C=C and X=X) bonds inside some fragments also deserves our attention here as an important particular case. As is seen from Eqs.(31)-(33), these dipoles originate from superposition of polarizations $p^{(1)long}_{(3)I-L}$ and $p^{(1)int}_{(3)I-L}$ and from the latter alone for fragments of the first and of the second group, respectively. As a result, positive and negative signs of total dipoles of C=C bonds (L) correspondingly follow for fragments I, VI and II, viz. these coincide with $3\alpha\gamma^2/16$, $2\alpha\gamma^2/16$ and $-\alpha\gamma^2/16$, respectively (Table 1). Thus, the left X=C and X=X bonds exert attractive effects upon electrons of their initially-homopolar neighbours (L), whereas the relevant effect of a C=X bond is of a repulsive nature. An analogous result easily follows also from comparison of secondary dipoles of X=X bonds inside fragments VII, VI and VIII.

It is seen, therefore, that total contributions of individual fragments I–L to secondary dipoles of the double bonds concerned (I and L) actually depend on the outcome of an interplay between $\alpha$-dependent indirect interactions $G^{(1)}_{(2)il}$ and $D^{(1+)}_{(2)il}$. An analogous state of things was observed also in respect of formation of respective contributions to the overall alterations in conjugation energies [40], viz. fragments of the first group and those of the second one proved to be characterized by positive (stabilizing) and negative (destabilizing) contributions, respectively. Moreover, a certain parallelism seems to exist between the longitudinal and the internal polarizations, on the one hand, and stabilizing and destabilizing components of the above-specified energetic contributions, on the other hand.

Finally, the overall pattern of the intrafragmental charge redistribution follows from superposition of the interbond charge transfer (represented by the partial populations $x^{(1),(+)l,(-)i}_{(3)}$) and of the above-discussed secondary dipoles ($d^{(1)}_{(3)I(L)}$ and $d^{(1)}_{(3)L(I)}$). It is then seen that both factors effect in the same direction and a significant shift of population from the right to the left consequently follows for fragments of the first group but not for those of the second one and this outcome causes little surprise. Indeed, an analogous qualitative distinction is known between 1- and 2-mono-substituted butadienes and hexatrienes in respect of the overall character of redistribution of $\pi$-electrons. In particular, the so-called global and local $\pi$-electron flow, respectively, is ascribed to these isomers [13, 41, 42] on the basis of application of the well-known curly arrow formalism (arrow pushing) [14-16]. Invoking of a certain extended hydrocarbon as a model of the original heteroatom-containing system [41, 42] serves to support this expectation. A more rigorous accounting for the same distinction between substituted isomers follows from graph-theoretical studies of extinction rates of the relevant atom-atom polarizabilities [13]. The above-obtained results provide us with an alternative interpretation of the same phenomenon in terms of local interactions of BOs.

## Conclusion

The results of the present study yield an image of the heteroatom influence in acyclic polyenes. The principal points here are as follows:

1. Behind the charge redistribution due to introduction of heteroatom(s), there are additional direct and indirect interactions of BOs of C=C bonds of the parent hydrocarbon. Moreover, one can imagine these interactions as undergoing a step-by-step expansion over the carbon backbone along with a simultaneous extinction of their relative magnitude: At the early stage, introduction of a heteroatom into a certain (say, the Ith) C=C bond of polyene is accompanied by emergence of direct intrabond self- and mutual interactions of BOs of only this bond. These newly-formed local interactions, in turn, potentially give birth to indirect interactions of a lower relative magnitude that embrace the nearest neighbourhood of the Ith bond (i.e. all two-membered conjugated fragments I–L, the Ith bond participates in) and so forth. Additivity both among these interactions and inside the latter (i.e. with respect to contributions of individual mediators) also is an essential aspect here.

2. The above-overviewed properties of extra interactions of BOs have their parallels in the relevant charge redistribution. This primarily refers to the high degree of additivity. Thus, the overall charge redistribution consists of two meaningful components, viz. of the pairwise charge transfer between formally-double bonds and of polarization of the latter (i.e. emergence of intrabond dipoles). Each of these components, in turn, is representable as superposition of contributions associated with more and more extended fragments of the molecule and characterized by a lower and lower significance. Local and transferable dipoles of individual X=C(C=X) bonds deserve mention here in the first place. Thereupon, additive and transferable contributions of separate two-membered conjugated fragments I–L follow, viz. the partial populations transferred between the Ith and Lth bonds along with secondary (induced) parallel and anti-parallel dipoles of these bonds, jointly referred to as the longitudinal and the internal polarization of the fragment I–L, respectively.

3. Separate local and semi-local contributions to the overall charge redistribution may be traced back to specific additional direct or indirect interactions of BOs of the parent polyene. In particular, the local (primary) dipole of an individual X=C (or C=X) bond originates from the newly-emerging direct intrabond interaction ($G^{(1)}_{(2)ii}$) between the BBO of the respective (Ith) C=C bond and its counterpart (ABO). Further, the partial population transferred inside a certain fragment I–L and the longitudinal polarization of the latter are determined by the indirect interaction ($G^{(1)}_{(2)il}$) between the BBO of the Ith bond and the ABO of the Lth one (or vice versa). Meanwhile, the relevant internal polarization is proportional to the analogous interaction between the two BBOs ($D^{(1+)}_{(2)il}$). This implies that contributions of the fragment I–L to total secondary (induced) dipoles of the Ith and Lth bonds depend on the outcome of an interplay between the two indirect intrafragmental interactions ($G^{(1)}_{(2)il}$ and $D^{(1+)}_{(2)il}$).

4. Analysis of expressions for the above-specified decisive intrafragmental interactions shows that the interaction $G^{(1)}_{(2)il}$ takes a non-zero and uniform value for some of fragments I–L (viz. I, III, VI and VII) and vanishes for the remaining ones (II, IV, V, VIII and IX) due to addition and cancellation of increments of individual mediators, respectively. Fragments of the first group are then characterized by both the partial transferred population and the longitudinal polarization of a significant magnitude that effect in the same direction and predominate over the internal polarization in addition. As a result, these fragments prove to be of a high ultimate polarity. Meanwhile, the above-specified two contributions vanish in the second case and the consequent overall polarities of fragments concerned either are negligible or coincide with zero. Moreover, the primary dipoles of separate X=C(C=X) bonds and their secondary (induced) counterparts are of opposite signs for fragments of the second group so that the observed polarities of these bonds also are relatively low. The above-concluded qualitative distinction between polarities of fragments of separate groups enriches the earlier-suggested classification of the same fragments on the basis of signs of the relevant contributions to the overall alterations in the conjugation energies [40].

## Methods

### *The direct way of derivation of the charge-bond order matrix*

Derivations of one-electron density matrices of molecules, in general, and of the relevant charge-bond order (CBO) matrices, in particular, usually are based on employment of respective molecular orbitals (MOs) represented in the form of linear combinations of atomic or bond orbitals (i.e. of the MO LCAO coefficients [18, 19, 43, 44]). The MOs, in turn, follow from solution of the standard eigenvalue equation for the respective Hamiltonian matrix. This way of derivation of the CBO matrix may be called the indirect one (i.e. via MOs).

It is much less known, however, that the same CBO matrix ($P$) is alternatively obtainable directly, i.e. without any reference to MOs [35]. To this end, the following system of matrix equations should be solved

$$[\bar{H}, P]_{-} = 0, \quad P^2 = 2P, \quad Tr\, P = 2n, \tag{40}$$

where $\bar{H}$ is the relevant Hamiltonian matrix, notations $[…,…]_{-}$ and $Tr$ stand for a commutator of matrices and for a trace of the latter, respectively, and $2n$ coincides with the total number of electrons in the system concerned. The first relation of Eq.(40) (the commutation condition) is the main physical requirement determining the matrix $P$ and resulting from the Dirac equation for the time-independent Hamiltonian [35]. Meanwhile, the remaining relations of Eq.(40) are additional system-structure-independent restrictions following from the idempotence requirement ($\Pi^2 = \Pi$) for the projector $\Pi = P/2$ and from the charge conservation condition, respectively.

An important advantage of the direct derivation of the CBO matrix over the indirect one consists in feasibility of results of a high degree of generality. Indeed, the standard eigenvalue equation is solvable for the Hamiltonian matrix of an individual molecule only and this also consequently refers to the subsequent obtaining of the matrix $P$. Meanwhile, solution of equations of Eq.(40) is performable for a whole class of related compounds, provided that the relevant common Hamiltonian matrix is easily constructable and/or known [32-34].

To illustrate this feasibility, let us dwell on molecules and/or molecular systems representable by two weakly-interacting well-separated subsets of basis functions $\{\phi_{(+)}\}$ and $\{\phi_{(-)}\}$ [34]. Further, let these subsets contain $n$ and $m$ orbitals, respectively, that are correspondingly double-occupied and vacant before perturbation. [Note that $n$ does not necessarily coincide with $m$]. The most general form of the relevant common Hamiltonian matrix then as follows

$$\bar{H} = \bar{H}_{(0)} + \bar{H}_{(1)} = \begin{vmatrix} E_{(+)} & 0 \\ 0 & -E_{(-)} \end{vmatrix} + \begin{vmatrix} T & R \\ R^* & Q \end{vmatrix}, \tag{41}$$

where $\bar{H}_{(0)}$ and $\bar{H}_{(1)}$ stand for zero- and first order members, respectively. Submatrices $E_{(+)} + T$ and $-E_{(-)} + Q$ correspondingly contain the Hamiltonian matrix elements between orbitals inside subsets $\{\phi_{(+)}\}$ and $\{\phi_{(-)}\}$, whereas $R$ embraces the intersubset interactions. Besides, intrasubset interactions of the zero order magnitude are allowed in the matrix $\bar{H}$ and these coincide with off-diagonal elements of submatrices $E_{(+)}$ and $E_{(-)}$. It also deserves emphasizing that neither the internal constitutions of submatrices nor their dimensions ($n$ and $m$) are specified in the matrix $\bar{H}$. Finally, comparison of the matrix $\bar{H}$ to $H'$ of Eq.(3) shows the latter to coincide with a particular case of the former, where $E_{(+)}=E_{(-)}=I$ and $n=m$. Thus, polyenes with heteroatom(s) offer us an example of systems representable by the matrix $\bar{H}$.

Solution of equations of Eq.(40) for our generalized Hamiltonian matrix $\bar{H}$ of Eq.(41) [32] was based on the following points: First, the matrix $P$ has been sought as a sum of increments $P_{(k)}$ of increasing orders ($k$) with respect to blocks of the first order Hamiltonian matrix ($T$, $Q$ and $R$), viz.

$$P = \sum_{k=0}^{\infty} P_{(k)} = P_{(0)} + P_{(1)} + P_{(2)} + ... \qquad (42)$$

Second, both the total matrix $P$ and its individual increments $P_{(k)}$, $k=0, 1, 2...$ have been initially represented via four submatrices (blocks) of appropriate dimensions (i.e. coinciding with $n \times n$, $n \times m$, etc). Consequently, separate blocks of the matrix $P$ also take the form of power series like that of Eq.(42). Finally, the above-specified overall form of the matrix $P$ has been substituted into relations of Eq.(40) and terms of each order ($k$) separately have been collected (as is usual in perturbative approaches). Let us now turn to the results.

The zero order member ($P_{(0)}$) of the CBO matrix $P$ follows straightforwardly from Eq.(40) after collecting terms of the zero order ($k=0$) and coincides with that of Eq.(6). Furthermore, the off-diagonal blocks of the subsequent corrections $P_{(k)}$, $k=1, 2...$ result from the commutation condition of Eq.(40). Thus, after introducing a notation $-2G_{(k)}$ for the off-diagonal block of the correction $P_{(k)}$, we obtain the following system of equations for matrices $G_{(k)}$, viz.

$$E_{(+)}G_{(k)} + G_{(k)}E_{(-)} + W_{(k)} = 0 , \qquad (43)$$

where

$$W_{(1)} = R, \qquad W_{(2)} = TG_{(1)} - G_{(1)}Q, \qquad (44)$$
$$W_{(3)} = TG_{(2)} - G_{(2)}Q - (RG_{(1)}^{*}G_{(1)} + G_{(1)}G_{(1)}^{*}R), etc.$$

As already mentioned, the particular case $E_{(+)}=E_{(-)}=I$ inside Eq.(41) corresponds to our polyenes with heteroatom(s). Matrix equations of Eq.(43) are then solvable algebraically and yield expressions for matrices $G_{(k)}$, $k=1,2...$ shown in Eq.(8). Meanwhile, the remaining blocks of corrections $P_{(k)}$ (taking diagonal positions inside the latter) result from the indempotence requirement of Eq.(40) and are consequently expressible algebraically via products of matrices $G_{(k)}$ of lower orders ($k$) as shown in Eq.(7).

The actual occupation numbers (populations) of individual basis orbitals follow from respective diagonal elements of submatrices taking diagonal positions within contributions $P_{(0)}$ and $P_{(k)}$, $k=1, 2...$ after summing them over the order parameter $k$ in accordance with Eq.(42). As a result, expressions of Eq.(9) are easily obtainable. At the same time, the intrabond bond orders of Eq.(13) follow from analogous sums over $k$ of respective diagonal elements of the remaining submatrices of the same contributions $P_{(0)}$ and $P_{(k)}$, $k=1, 2...$ taking the off-diagonal positions inside the latter.